\documentclass{ifacconf}

\usepackage{graphicx}      %
\usepackage{natbib}        %
\usepackage{amsmath,amssymb,amsfonts}
\usepackage{mathrsfs}
\usepackage{booktabs}
\usepackage{bibentry}

\usepackage{tikz}
\usepackage{xspace}

\DeclareRobustCommand{\legendline}[1]{\hspace{-2pt}
\tikz[#1,line width=1.5pt,baseline=-0.5ex]{\draw (0,0) -- (.35,0);}
\hspace{-2pt}}

\definecolor{mblue}{rgb}{0,0.4470,0.7410}
\definecolor{morange}{rgb}{0.8500,0.3250,0.0980}
\definecolor{myellow}{rgb}{0.9290,0.6940,0.1250}
\definecolor{mpurple}{rgb}{0.4940,0.1840,0.5560}
\definecolor{mgreen}{rgb}{0.4660,0.6740,0.1880}
\definecolor{mcyan}{rgb}{0.3010,0.7450,0.9330}
\definecolor{mred}{rgb}{0.6350,0.0780,0.1840}
\definecolor{mgreenblue}{rgb}{0.0,1.0,0.5}
\definecolor{parulablue}{rgb}{0.2431,0.1490,0.6588}
\definecolor{parulalblue}{RGB}{39,151,235}
\definecolor{parulagreen}{RGB}{129,204,89}
\definecolor{parulayellow}{RGB}{249,251,21}
\definecolor{wintergreen}{cmyk}{0.61,0,0.74,0}

\newcommand{\scalemath}[2]{\scalebox{#1}{\mbox{\ensuremath{\displaystyle #2}}}}

\newcommand{\sn}[1]{\ensuremath{\cdot 10^{#1}}\xspace} 

\DeclareFontFamily{OT1}{pzc}{}
\DeclareFontShape{OT1}{pzc}{m}{it}{ <-> s*[1.1] pzcmi7t }{}
\DeclareMathAlphabet{\mathpzc}{OT1}{pzc}{m}{it}

\newcommand{\customurl}[2]{\href{#1}{\path{#2}}}

\newcommand{\mc}[1]{\mathcal{#1}}

\newcommand{\mr}[1]{\mathrm{#1}}

\newcommand{\mb}[1]{\mathbb{#1}}

\newcommand{\C}[1]{\ensuremath{\mc{C}_{#1}}\xspace}   %

\newcommand{\reals}{{\mb{R}}}

\newcommand{\integ}{\ensuremath{\mb{Z}}\xspace}
 \newcommand{\nninteg}{\ensuremath{\integ_0^+}\xspace}   %

\DeclareMathOperator{\col}{col}
\DeclareMathOperator{\diag}{diag}
\DeclareMathOperator{\vect}{vec}

\newcommand{\norm}[1]{\left\lVert#1\right\rVert}

\newcommand{\otherTraj}[1]{\expandafter\tilde #1}

\newcommand{\parTraj}[1]{\expandafter\bar #1}

\newcommand{\citePete}{(Annoni et al., 2017)\xspace} %

\makeatletter
\let\old@ssect\@ssect %
\makeatother

\usepackage{hyperref}

\makeatletter
\def\@ssect#1#2#3#4#5#6{%
  \NR@gettitle{#6}%
  \old@ssect{#1}{#2}{#3}{#4}{#5}{#6}%
}
\makeatother
\let\oldcitep\citep
\renewcommand{\citep}[1]{\begin{NoHyper}\oldcitep{#1}\end{NoHyper}}

\makeatletter
\g@addto@macro\normalsize{
  \setlength\abovedisplayskip{.2em}
	\setlength\belowdisplayskip{.2em}
	\setlength\abovedisplayshortskip{.2em}
	\setlength\belowdisplayshortskip{.2em}
	\setlength\parskip{.1em}
}

\begin{document}
\begin{frontmatter}

\title{Learning Reduced-Order Linear Parameter-Varying Models of Nonlinear Systems\thanksref{footnoteinfo}} 

\thanks[footnoteinfo]{This work has been supported by Sioux Technologies B.V., The MathWorks Inc., and by the European Union within the framework of the National Laboratory for Autonomous Systems (RRF-2.3.1-21-2022-00002). Opinions, findings, conclusions or recommendations expressed in this paper are those of the authors and do not necessarily reflect the views of Sioux Technologies B.V., the MathWorks Inc., or the European Union.}

\author[mathware]{Patrick J. W. Koelewijn}
\author[mathworks]{Rajiv Singh}
\author[umich]{Peter Seiler}
\author[tue,sztaki]{Roland T\'oth} 

\address[mathware]{Sioux Technologies B.V., Eindhoven, The Netherlands. (e-mail: research@patrickkoelewijn.nl)}
\address[tue]{Control Systems Group, Eindhoven University of Technology, Eindhoven, The Netherlands (e-mail: r.toth@tue.nl)}
\address[mathworks]{The MathWorks, Inc., Natick, USA, (e-mail: rsingh@mathworks.com)}
\address[umich]{Department of Electrical Engineering and
Computer Science, University of Michigan, Ann Arbor, USA, (e-mail: pseiler@umich.edu)}
\address[sztaki]{Systems and Control Laboratory, HUN-REN Institute for Computer Science and Control, Budapest, Hungary}

\begin{abstract}                %
In this paper, we consider the learning of a \emph{Reduced-Order Linear Parameter-Varying Model} (ROLPVM) of a nonlinear dynamical system based on data. This is achieved by a two-step procedure. In the first step, we learn a projection to a lower dimensional state-space. In step two, an LPV model is learned on the reduced-order state-space using a novel, efficient parameterization in terms of neural networks. The improved modeling accuracy of the method compared to an existing method is demonstrated by simulation examples. \vspace{-3mm}
\end{abstract}


\end{frontmatter}

\section{Introduction}
Across numerous industries in engineering, there is the constant push to improve the performance of dynamic systems. To realize these improvements, the use of large-scale, nonlinear, physics-based models for accurately modeling the dynamics of these systems has become common in the industry. These models are then used to predict, analyze, and/or regulate the dynamical behavior of these systems. However, this traditional approach is becoming  more and more difficult and even impossible with the increase in complexity and size of these systems.

Therefore, obtaining \emph{Reduced-Order Models} (ROMs) with acceptable accuracy has seen great interest over the last years.  With reduced-order modeling, the goal is to capture the dynamics of interest in a model with low complexity, e.g., a low state dimension for dynamical systems in state-space representation, but with adequate accuracy. Often, this is also combined with capturing the dynamics in a model with an easier-to-work-with structure, such as in a (set of) \emph{Linear Time-Invariant} (LTI) model(s) or \emph{Linear-Parameter Varying} (LPV) models. 

Various data-driven methods exist for reduced-order modeling such as %
\emph{Sparse Identification of Nonlinear Dynamics} (SINDy) \citep{Brunton2016}, Neural ODEs  \citep{Chen2018}, or \emph{Local Linear Model Trees} (LOLIMOT) \citep{Nelles2000}, to name a few. However, the downside of these methods is that while the resulting model is often lower dimensional, benefiting simulation speed, their structure is generally still nonlinear, which makes using the resulting model for analysis or control rather difficult. The method in \citePete, improves upon this by learning a \emph{Reduced-Order LPV Model} (ROLPVM). The benefit of learning a ROLPVM is that there exist many methods \citep{Apkarian1995,Cox2018a,DeCaigny2012}  and software tools \citep{DenBoef2021} to analyze and design controllers for LPV models. However, the downside of %
\citePete is that it uses local data around equilibrium points of the system to obtain local LTI models, interpolated to construct an LPV model. Due to the reliance on local data, the full nonlinear behavior of the underlying system might not be captured sufficiently well, limiting model accuracy.

Compared to LPV conversion methods that keep the original state dimension, LPV identification methods \citep{Verhoek2022}, \citep{Masti2021} can also be used to obtain a reduced-order LPV model based on input-output data only. However, they cannot explicitly exploit that for reduced-order modeling, knowledge of the state is generally available nor can they explicitly characterize the reduction transformation w.r.t. the original state. 

In this work, we propose a data-driven method that learns a global ROLPVM of a nonlinear dynamical system. The approach, which can be seen as a fusion of (Annoni et al., 2017; Koelewijn and T\'oth, 2020),  is based on a novel neural network architecture,  able to efficiently learn a ROLPVM from simulation or experimental data of the system. To do so, we extend the neural network representation for LPV scheduling reduction in \citep{Koelewijn2020a} to allow for modeling of full LPV state-space representations, while we extend the state reduction approach in \citePete from a linear static state-transformation to a nonlinear one by the help of deep learning. Compared to other reduced-order modeling methods, we (i) learn a structured reduced-order model in the form of an LPV model, which is beneficial if the model is also to be used for analysis and controller design, and (ii) learn a global ROLPVM based on simulation/experimental data of the underlying nonlinear system, instead of using local data, which allows for improved modeling accuracy.

\subsection*{Notation}
The set of real numbers is denoted by $\reals$. The set of non-negative integers is denoted by $\nninteg$. %
For a vector $x=\begin{bmatrix}x_1&\dots&x_n\end{bmatrix}$, $\diag(x)$ denotes a diagonal matrix with the scalar values $x_1,\dots,x_n$ on the diagonal. The vector $\begin{bmatrix}
	x_1^\top & \cdots & x_n^\top
\end{bmatrix}^\top$ is denoted by $\col(x_1, \dots, x_n)$. The column-wise vectorization of a matrix $M$ is given by $\vect(M)$. The weighted 2-norm of a vector $x$ is denoted by $\norm{x}_{2,W} = \sqrt{x^\top W x}$ and $\norm{x}_2 = \sqrt{x^\top x}$ for $W=I$. %
\vspace{-0.5mm}
\section{Problem Statement}\label{sec:problem} \vspace{-0.5mm}
We consider \emph{Discrete-Time} (DT) data-generating nonlinear systems in state-space form given by 
\begin{subequations}\label{eq:nlsys}
	\begin{align}
		x(k+1) &= f(x(k),u(k)) + w(k);\\
		y(k) &= h(x(k), u(k))+e(k);
	\end{align}
\end{subequations}
where $k\in\nninteg$ is the discrete time,  $x(k)\in\reals^{n_\mr{x}}$ is the state, $u(k) \in\reals^{n_\mr{u}}$ is the input to the system, $y(k)\in\reals^{n_\mr{y}}$ is the output of the system, and $w(k)\in\reals^{n_\mr{x}}$ and $e(k)\in\reals^{n_\mr{y}}$ are realizations of i.i.d. white noise processes with finite variances $\Sigma_\mr{w}\in\reals^{n_\mr{x}\times n_\mr{x}}$ and $\Sigma_\mr{e}\in\reals^{n_\mr{y}\times n_\mr{y}}$, respectively. If our objective is data-driven reduction based on simulation data, then $w(k)$ and $e(k)$ are zero. On the other hand, if our objective is reduced-order model identification based on experimental data, then $w(k)$ and $e(k)$ are \emph{not} zero. %

For the system given by \eqref{eq:nlsys}, we assume we have gathered data in the form of the data sets:
\begin{equation}
\begin{alignedat}{3}
		\mb{X} &= \left\lbrace x(k)\right\rbrace_{k=0}^{N-1}, &&\quad \mb{X}_+ &&= \left\lbrace x(k+1)\right\rbrace_{k=0}^{N-1},\\
		\mb{U} &= \left\lbrace u(k)\right\rbrace_{k=0}^{N-1}, &&\quad \mb{Y} &&= \left\lbrace y(k)\right\rbrace_{k=0}^{N-1},
\end{alignedat}
\end{equation}
i.e., $N+1$ state data points, and $N$ input and output data points. The collection of these data sets is denoted by:
\begin{equation}
	\mc{D} =\left\lbrace \left(x(k), x(k+1), u(k), y(k)\right) \right\rbrace_{k=0}^{N-1} .
\end{equation}
Note that to be able to capture the full behavior of \eqref{eq:nlsys}, the input excitation/data, $\mb{U}$, persistently exciting  w.r.t to a given region of interest in $\mb{X}$ and $\mb{U}$. However, for general nonlinear systems there are limited results available to ensure such properties \cite{Schoukens2019}. Further details on this is therefore outside of the scope of the paper. Therefore, we assume in this paper that $\mc{D}$ sufficiently covers the operating region deemed of interest by the user.

As an objective, we want to learn a ROLPVM that approximates the behavior of \eqref{eq:nlsys} for a given data set $\mc{D}$. We consider the ROLPVM in the state-space form:
\begin{subequations}\label{eq:rolpvm}\begin{equation}\label{eq:lpv}
\begin{bmatrix}
	z(k+1)\\y(k)
\end{bmatrix} = \underbrace{\begin{bmatrix}
	A(p(k)) & B(p(k))\\C(p(k)) & D(p(k))
\end{bmatrix}}_{M(p(k))}\begin{bmatrix}
	z(k)\\u(k)
\end{bmatrix}+\begin{bmatrix}
	z_\mr{o}\\y_\mr{o}
\end{bmatrix} ,
\end{equation}
\begin{equation}\label{eq:schedmap}
	p(k) = \mu(z(k), u(k)),
\end{equation} \end{subequations}
where $p(k) \in \reals^{n_\mr{p}}$ is the scheduling-variable, $z(k)\in\reals^{n_\mr{z}}$ is the reduced-order state with $n_\mr{z} < n_\mr{x}$. Moreover, $\mu:\reals^{n_\mr{z}}\times\reals^{n_\mr{u}}$ is the scheduling map, $A:\reals^{n_\mr{p}}\to\reals^{n_\mr{z}\times n_\mr{z}}$, $B:\reals^{n_\mr{p}}\to\reals^{n_\mr{z}\times n_\mr{u}}$, $C:\reals^{n_\mr{p}}\to\reals^{n_\mr{y}\times n_\mr{z}}$, and $D:\reals^{n_\mr{p}}\to\reals^{n_\mr{y}\times n_\mr{u}}$ are  matrix functions, and $z_\mr{o}\in\reals^{n_\mr{z}}$ and $y_\mr{o}\in\reals^{n_\mr{y}}$ are constant offset/bias terms.
Furthermore, we aim at learning a state-projection map $\psi:\reals^{n_\mathbf{x}} \to \reals^{n_\mathbf{z}}$, which maps from the full state $x$ to the reduced state $z$, and an inverse map $\psi^{\dagger}:\reals^{n_\mathbf{z}} \to \reals^{n_\mathbf{x}}$, which (approximately) maps $z$ back to $x$.
More concretely, in terms of objective, we want to learn a ROLPVM \eqref{eq:rolpvm} such that we minimize 
\begin{equation}\label{eq:romobjective}
	\min J(\theta) %
\end{equation}
where %
the cost $J$ is considered as
\begin{equation}\label{eq:J1}
	J(\theta) =  \frac{1}{N}\sum_{k=0}^{N-1}\norm{\begin{bmatrix}
		x(k+1)\\y(k)
	\end{bmatrix} - \begin{bmatrix}
		\hat{f}_\theta(x(k),u(k))\\\hat{h}_\theta(x(k),u(k))
	\end{bmatrix}}^2_{2,\Gamma}
\end{equation}
where $\Gamma\in\reals^{(n_\mr{x}+n_\mr{y})\times (n_\mr{x}+n_\mr{y})}$ is a weight matrix, $(x(k),\linebreak {x(k+1)}, u(k), y(k))\in\mc{D}$ and
\begin{subequations}
	\begin{align}
			\hat{f}_\theta(x,u) &= \psi^{\dagger}\bigl(A(p)\psi(x)+B(p)u\bigr),\\
			\hat{h}_\theta(x,u) &= C(p)\psi(x)+D(p)u,
	\end{align} 
\end{subequations}
with $p = \mu(\psi(x),u)$ and where $\theta$ are the parameters that define the ROLPVM \eqref{eq:rolpvm}, and the state-projections $\psi$ and $\psi^\dagger$. 

This means that our objective in terms of \eqref{eq:romobjective} is to minimize the \emph{Mean Squared Error} (MSE) of the one-step ahead state and output prediction\footnote{$J$ can easily be extended to also support $n$-step ahead prediction.} of our ROLPVM \eqref{eq:rolpvm} on the data set $\mc{D}$. %
Note that if $x(k)$ is not directly measurable, one can always use an extended (non-minimal) state-construction $x(k)=[\ y(k-1) \ \cdots \ y(k-n) \ u(k-1) \ \cdots\ u(k-n)\ ] $ with $n$ being large enough (greater than the lag of the system). 
In the next section, we will discuss our proposed method to achieve this objective.

\section{Method}\label{sec:method}\vspace{-2.5mm}
\subsection{Overview}\vspace{-1mm}
To achieve our objective of learning a ROLPVM of the form \eqref{eq:rolpvm} along with a state-projection map $\psi$ and the corresponding inverse map $\psi^{\dagger}$, we propose the following two-step procedure:
\begin{enumerate}
	\item We first learn a nonlinear state-projection $\psi$ and the corresponding inverse map $\psi^{\dagger}$ based on the state data $\mb{X}$, part of $\mc{D}$, such that the cost  
 \begin{equation}\label{eq:J2}
	J_1(\theta) = \frac{1}{N} \sum_{k=0}^{N-1} \norm{x(k)-\psi^\dagger(\psi(x(k)))}_{2,\Gamma_1},
\end{equation}
 is minimized, where $\Gamma_1 \in \reals^{n_\mr{x}\times n_\mr{x}}$ is a weight matrix. 
	\item Based on the learned state-projection $\psi$, we construct new data sets of the reduced-order state:
	\begin{equation}\label{eq:zdatamat}
	\begin{aligned}
		\mb{Z} &= \left\lbrace z(k)\right\rbrace_{k=0}^{N-1} = \left\lbrace \psi(x(k))\right\rbrace_{k=0}^{N-1}\!,\\
		\mb{Z}_+ &= \left\lbrace z(k+1)\right\rbrace_{k=0}^{N-1} =\left\lbrace \psi(x(k+1)) \right\rbrace_{k=0}^{N-1},
	\end{aligned}
	\end{equation}
	also resulting in %
	\begin{equation}\label{eq:zdataset}
	\mc{D}_\mr{z} \! = \! \left\lbrace \left(z(k), z(k+1), u(k), y(k)\right) \right\rbrace_{k=0}^{N-1}.
\end{equation}
	We then minimize
	\begin{multline}\label{eq:romreducedobjective}
	J_2(\theta) = \frac{1}{N}\sum_{k=0}^{N-1}\bigg\lVert\begin{bmatrix}
		z(k+1)\\y(k)
	\end{bmatrix} - \\[-1mm] \left(\! M(\mu(z(k),u(k)))	\!\begin{bmatrix}
		z(k)\\ u(k)
	\end{bmatrix}\!+\!\begin{bmatrix}
	z_\mr{o}\\y_\mr{o}
\end{bmatrix}\right)\bigg\rVert^2_{2,\Gamma_2},
\end{multline}
where $(z(k), z(k+1), u(k), y(k))\in\mc{D}_\mr{z}$. 
\end{enumerate}
Therefore, we solve the minimization of $J$ in a hierarchical sense \citep{Khandelwal2020}, by first minimizing $J_1$ in the first step and then optimizing {$J_2$ in} \eqref{eq:romreducedobjective}.

In \citePete, a comparable method is proposed to obtain a ROLPVM, where first also a state-projection is computed, which is then used to fit the LPV state-space matrices. However, there are two major advantages of our method compared to the one in \citePete. Namely, the state-projection in \citePete uses a linear PCA based projection, while the approach we propose, as we will show it later, uses a nonlinear state-projection. This will result in an improved reconstruction of the full order state $x$, thereby reducing the modeling error. Secondly, the approach in \citePete uses local data to construct local models while our approach uses global simulation/experimental data  to directly construct a global LPV model. This simplifies the procedure, as the method \citePete requires obtaining local data at many operating points, while our approach only requires a single data set. Moreover, as the method in \citePete only uses local data, it might not be able to adequately capture all the nonlinear behavior of the system, while our approach uses global data of the system, and hence can do so. 

In the following subsections, we will give more details on each of the individual steps of our proposed procedure. First, we will provide details about step one, i.e., how to learn the state-projection.

\subsection{Learning the state-projection}\label{sec:stateproj}
As aforementioned, in step one of our procedure, we are interested in learning the state-projection (encoder function) $\psi:\reals^{n_\mathbf{x}} \to \reals^{n_\mathbf{z}}$ and the inverse projection (decoder function) $\psi^{\dagger}:\reals^{n_\mathbf{z}} \to \reals^{n_\mathbf{x}}$ such that $x \approx \psi^{\dagger}(\psi(x))$ for all $x\in\mb{X}$. This corresponds to a classical dimensionality reduction problem for which there exists a variety of techniques such as  \emph{Principal Component Analysis} (PCA), \emph{Kernel PCA} (KPCA) \citep{Schollkopf1998}, or \emph{Autoencoders} (AEs) \citep{Kramer1991}.

\begin{figure}
	\centering
	\includegraphics[scale=0.8]{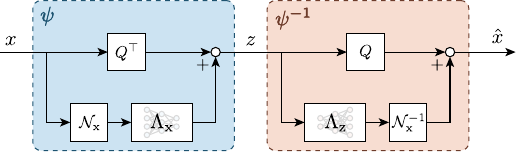}
	\vspace{-1ex}
	\caption{The architecture for state-projection.}
	\label{fig:stateproj}
\end{figure}

In this paper, we propose a combined approach using linear PCA augmented by a nonlinear AE to perform the state-projection. The architecture of the proposed state-projection is depicted in Fig. \ref{fig:stateproj}, where $Q\in\reals^{n_\mr{x}\times n_\mr{z}}$ is a semi-orthogonal matrix\footnote{A matrix $Q\in\reals^{n\times m}$ with $m \leq n$ is semi-orthogonal if $Q^\top Q = I$.}
 corresponding to the projection matrix of linear PCA, $\Lambda_\mr{x}:\reals^{n_\mr{x}}\to\reals^{n_\mr{z}}$ is an AE type encoder network, $\Lambda_\mr{z}:\reals^{n_\mr{z}}\to\reals^{n_\mr{x}}$ is an AE type decoder network, and $\mc{N}_\mr{x}:\reals^{n_\mr{x}}\to\reals^{n_\mr{x}}$ and $\mc{N}_\mr{x}^{-1}:\reals^{n_\mr{x}}\to\reals^{n_\mr{x}}$ are normalization and inverse normalization functions, respectively. 

More specifically, for the proposed state-projection architecture, we have that
\begin{subequations}\label{eq:stateprojboth}
\begin{equation}\label{eq:stateproj}
	\psi(x) = Q^\top x + \Lambda_\mr{x}(\mc{N}_\mr{x}(x)),
\end{equation}
and 
\begin{equation}\label{eq:stateprojinv}
	\psi^{\dagger}(z) = Qz + \mc{N}_\mr{x}^{-1}(\Lambda_\mr{z}(z)).
\end{equation}	
\end{subequations}
Here, the linear PCA projection matrix $Q$ is computed based on our data set $\mb{X}$. This is done as follows. Introduce the matrix
\begin{equation}
	X = \begin{bmatrix}
		x(0) & \cdots & x(N-1)
	\end{bmatrix} \in \reals^{n_\mr{x} \times N},
\end{equation}
containing the data in $\mb{X}$.
Using a \emph{Singular Value Decomposition} (SVD), we then factorize $X$ as follows
\begin{equation}
	X = \begin{bmatrix}
		U_\mr{z} & U_\mr{m}
	\end{bmatrix}\begin{bmatrix}
		S_\mr{z} & 0 & 0\\0 & S_\mr{m} & 0
	\end{bmatrix}\begin{bmatrix}
		V_\mr{z} & V_\mr{m}
	\end{bmatrix}^\top,
\end{equation}
where $S_\mr{z}\in\reals^{n_\mr{z}\times n_\mr{z}}$ is a diagonal matrix containing the first $n_\mr{z}$ largest singular values, with $S_\mr{m}$ containing the remaining $n_\mr{x} - n_\mr{z}$ singular values. For this partitioning of ${X}$, our projection matrix $Q$ is then equal to $U_\mr{z}\in\reals^{n_\mr{x}\times n_\mr{z}}$, i.e., $Q = U_\mr{z}$. Here, $n_\mr{z}$ determines the reduced state dimension, which can be chosen by examining the relative magnitude of the singular values of $X$.

Next, we define $\Lambda_\mr{x}$ and $\Lambda_\mr{z}$ in \eqref{eq:stateprojboth} as \emph{(Deep) Neural Networks (DNNs)} with $n$ layers of the form
\begin{subequations}\label{eq:dnn}
\begin{align}
	\vartheta_1 &= \sigma_1(W_1 \vartheta_{0} + b_1),\\
	\vartheta_i &= \sigma_i(W_i \vartheta_{i-1} + b_i)\quad \text{for $i \in \mathbb{I}_2^{n -1}$},\\
	\vartheta_n &= W_{n}\vartheta_{n-1} + b_{n},
\end{align}
\end{subequations}
where $\mathbb{I}_{\tau_1}^{\tau_2}=\{i\in\mathbb{N} \mid \tau_1\leq i \leq \tau_2\}$ is an index set, $\vartheta_i\in\reals^{n_{\mr{\vartheta},i}}$ for $i\in\mathbb{I}_0^{n}$  are the values of the neurons of layer $i$ with $i=0$ corresponding to the input layer and  $i=n$ to the output layer, $W_i\in\reals^{n_{\mr{\vartheta},i}\times n_{\mr{\vartheta},i-1}}$ and $b_i\in\reals^{n_{\mr{\vartheta},i}}$ %
are the weight matrices and bias terms of the layers, and $\sigma_i:\reals^{n_{\mr{\vartheta},i}}\to \reals^{n_{\mr{\vartheta},i}}$ is an array of activation functions forming layer $i$. In case of the encoder $\Lambda_\mr{x}$, $\vartheta_0 = \mc{N}_\mr{x}(x)\in\reals^{n_\mr{x}}$ and $\vartheta_n \in\reals^{n_\mr{z}}$, while for the decoder $\Lambda_\mr{z}$, $\vartheta_0 = z\in\reals^{n_\mr{z}}$ and $\vartheta_n \in\reals^{n_\mr{x}}$. Later, we will discuss how we train the weights and biases of $\Lambda_\mr{x}$ and $\Lambda_\mr{z}$.

The normalization function $\mc{N}_\mr{x}$ in \eqref{eq:stateprojboth} is given by
\begin{equation}\label{eq:normalizeX}
	\mc{N}_\mr{x} = N_\mr{x}\cdot (x-x_\mr{mean}),
\end{equation}
where $x_\mr{mean} = \frac{1}{N}\sum_{k=0}^{N-1} x(k)\in\reals^{n_\mr{x}}$ is the mean of the data in $\mb{X}$ and
\begin{equation}
N_\mr{x} = \diag\left(
\norm{x}_\infty
\right )^{-1} \in\reals^{n_\mr{x}\times n_\mr{x}},
\end{equation}
where $\norm{\cdot}_\infty$ denotes the \emph{element-wise} $\ell_\infty$ norm, i.e., the element-wise maximum absolute value over the elements in $\mb{X}$. Hence, $\mc{N}_\mr{x}$ corresponds to normalizing $x\in\mathbb{X}$ to $[-1,\, 1]^{n_\mr{x}}$ and centering it to zero. Based on the definition of $\mc{N}_\mr{x}$ in \eqref{eq:normalizeX}, it follows that
\begin{equation}
	\mc{N}_\mr{x}^{-1}(\bar{x}) = N_\mr{x}^{-1}\bar{x}+x_\mr{mean}.
\end{equation}

To train the weights and biases of the DNNs $\Lambda_\mr{x}$ and $\Lambda_\mr{z}$, we first compute the linear PCA projection matrix $Q$ and the normalization functions $\mc{N}_\mr{x}$ and $\mc{N}_\mr{x}^{-1}$ based on our data set $\mb{X}$, which are then fixed in the architecture depicted in Fig. \ref{fig:stateproj}. We then train $\Lambda_\mr{x}$ and $\Lambda_\mr{z}$ with the architecture depicted in Fig. \ref{fig:stateproj} on the state data in $\mb{X}$ such that the cost $J_1$ \eqref{eq:J2} is minimized where $\psi$ and $\psi^{\dagger}$ are given in \eqref{eq:stateprojboth}. Note that due to the normalization of the data, a natural choice of weighting in \eqref{eq:J2} is $\Gamma_1=\sqrt{N_\mr{x}}$, but different weighting can also be applied by the user.

Similar to standard machine learning methods involving (D)NNs, we use a \emph{Stochastic Gradient Descent} (SGD) method, such as Adam \citep{Kingma2014}, to minimize \eqref{eq:J2} and learn $\Lambda_\mr{x}$ and $\Lambda_\mr{z}$. After the construction and learning of $\psi$ as we previously described, we use it for the next step in our procedure to learn the ROLPVM of the form \eqref{eq:rolpvm}.

\subsection{Learning the ROLPVM}\label{sec:lpvnn-learning}

For step two of our approach, we aim to optimize \eqref{eq:romreducedobjective} to learn a ROLPVM of the form \eqref{eq:rolpvm}, represented by the state-space matrix functions forming $M$, a scheduling-map $\mu$, and offset terms $z_\mr{o}$ and $y_\mr{o}$. For our proposed method, we will represent both the matrix function $M$ and the scheduling-map $\mu$ by (D)NNs.

\vspace{1mm}
{\bf Training on the state-difference:}
While we could directly learn $M$, $\mu$, $z_\mr{o}$, and $y_\mr{o}$ such that we minimize our objective \eqref{eq:romreducedobjective} on the data set $\mc{D}_\mr{z}$, doing so could be problematic if the differences between $z$ and $z_+$ are small, e.g., in case of the data corresponds to sampling of a continuous-time system with a small sampling time. Therefore, we first transform the mapping and data before training, namely, we consider learning the mapping:
\begin{equation}\label{eq:mdeltamap}
		\begin{bmatrix}
		z_\Delta\\ y
	\end{bmatrix} = M_\Delta(\mu(z,u))\begin{bmatrix}
		z\\ u
	\end{bmatrix}+\begin{bmatrix}
	z_\mr{o}\\y_\mr{o}
\end{bmatrix} ,
\end{equation}
where $z_\Delta(k)$ corresponds to $z(k+1) - z(k)$, i.e., we learn the mapping to the time-difference of the reduced-order state. The relation between $M$ and $M_\Delta$ is then given by: 
\begin{equation}\label{eq:mdeltarelation}
	M_\Delta(p) = M(p) - \begin{bmatrix}
		I & 0\\0 & 0
	\end{bmatrix} = \begin{bmatrix}
	A(p) - I & B(p)\\C(p) & D(p)
\end{bmatrix}.
\end{equation}
Consistent with this, we also construct the data set
\begin{equation}
	\mb{Z}_\Delta = \left\lbrace z_\Delta(k)\right\rbrace_{k=0}^{N-1}= \left\lbrace z(k+1)-z(k)\right\rbrace_{k=0}^{N-1}.\\
\end{equation}

\vspace{1mm}
{\bf Model parameterization:}
In this paper, we restrict the ROLPVM to have an affine scheduling dependency. Therefore, we parameterize $M_\Delta(p)$ as follows:
\begin{equation}\label{eq:Maffine}
	M_\Delta(p) = M_{\Delta,0} + \sum_{i=1}^{n_\mr{p}} M_{\Delta,i}\, p_i,
\end{equation}
where $M_{\Delta,i} \in \reals^{(n_\mr{z}+n_\mr{y})\times (n_\mr{z}+n_\mr{u})}$  with $i \in \mathbb{I}_0 ^{n_\mr{p}}$ and $p_i$ is the $i^\mathrm{th}$ element of $p$, which is equal to the $i^\mathrm{th}$ element of the scheduling-map $\mu$. Note that this will also result in $M$ having an affine scheduling dependency through the relation in \eqref{eq:mdeltarelation}. While other scheduling dependencies can also be considered, such as rational or even generic nonlinear scheduling dependencies, considering an affine scheduling dependency has several benefits. Namely, 1) avoids complexity in the learning of a composite function composed from the scheduling dependency and the scheduling-map $\mu$; %
2) there exist many efficient methods for stability and performance analysis and controller synthesis w.r.t. LPV models with affine scheduling dependency \citep{Apkarian1995,Cox2018a,DeCaigny2012}. This allows us to efficiently analyze and design controllers for the nonlinear system described by the learned ROLPVM. 

Inspired by the method in \citep{Koelewijn2020a} for scheduling dimension reduction, we can represent the affine matrix function $M_\Delta$, given by \eqref{eq:Maffine}, by a linear single layer neural network. This is done as follows:
\begin{align}
	M_{\Delta,\mr{v}}(p) &= \vect(M_\Delta(p)), \\
	&= \underbrace{\begin{bmatrix}
		\vect(M_{\Delta,1}) & \dots & \vect(M_{\Delta,n_\mr{p}})
	\end{bmatrix}}_{W_\Delta} p + \underbrace{\vect(M_{\Delta,0})}_{b_\Delta}.\notag
\end{align}
$W_\Delta\in\reals^{((n_\mr{z}+n_\mr{y})\cdot (n_\mr{z}+n_\mr{u}))\times n_\mr{p}}$ and $b_\Delta \in \reals ^{((n_\mr{z}+n_\mr{y})\cdot (n_\mr{z}+n_\mr{u}))}$ are both parameters that we learn during training. The value of $M_\Delta(p)\in \reals^{(n_\mr{z}+n_\mr{y})\times (n_\mr{z}+n_\mr{u})}$ for a fixed value of $p$ can then be obtained by reshaping $M_{\Delta,\mr{v}}(p)\in \reals ^{((n_\mr{z}+n_\mr{y})\cdot (n_\mr{z}+n_\mr{u}))}$ back to a matrix in $\reals^{(n_\mr{z}+n_\mr{y})\times(n_\mr{z}+n_\mr{u})}$. 

\begin{figure}
	\centering
	\includegraphics{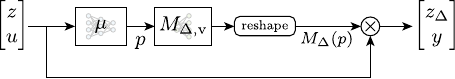}
	\caption{LPV-NN architecture for learning the ROLPVM.} 
	\label{fig:lpvann}
\end{figure}

As aforementioned, we also represent the scheduling-map $\mu$ by a DNN of the form \eqref{eq:dnn} where the number of layers, and type of activation functions are hyper-parameters, while the output size, $n_\mathrm{p}$, i.e., the scheduling dimension, is a user chosen complexity parameter. Accordingly, the input of %
the scheduling-map $\mu$ is %
$\col(z,u)$, %
while the
output %
is $p$. %
The full architecture to represent \eqref{eq:mdeltamap} is depicted in Fig. \ref{fig:lpvann}, which we will refer to it as the \emph{LPV Neural Network} (LPV-NN) architecture. 

\vspace{1mm}
{\bf Data normalization:}
For training of $M$, $\mu$, $z_\mr{o}$, and $y_\mr{o}$, we normalize our %
data sets $\mb{Z}_\Delta$, $\mb{Z}$, $\mb{U}$ and $\mb{Y}$ as follows:
\begin{equation}
	\bar{\mb{Z}}_\Delta = \left\lbrace \bar z_\Delta(k)\right\rbrace_{k=0}^{N-1} = \left\lbrace N_\mr{z_\Delta}z_\Delta(k)\right\rbrace_{k=0}^{N-1},
\end{equation}
where $z_\Delta(k) \in \mb{Z}_\Delta$ and 
\begin{equation}\label{eq:zdeltascaling}
N_\mr{z_\Delta} = \diag\left(
\norm{z_\Delta}_\infty
\right )^{-1} \in\reals^{n_\mr{z}\times n_\mr{z}},
\end{equation}
i.e., we normalize the data in $\mb{Z}_\Delta$ to lie in the interval $[-1,\,1]^{n_\mr{z}}$. Similarly, 
	$\bar{\mb{Z}} = \left\lbrace \bar z(k)\right\rbrace_{k=0}^{N-1}= \left\lbrace N_\mr{z}z(k)\right\rbrace_{k=0}^{N-1}$, $
	\bar{\mb{U}} = \left\lbrace \bar u(k)\right\rbrace_{k=0}^{N-1} = \left\lbrace N_\mr{u}u(k)\right\rbrace_{k=0}^{N-1}$, and $
	\bar{\mb{Y}} = \left\lbrace \bar y(k)\right\rbrace_{k=0}^{N-1} = \left\lbrace N_\mr{y}y(k)\right\rbrace_{k=0}^{N-1}$, where $N_\mr{z}$, $N_\mr{u}$, and $N_\mr{y}$ are similarly defined as \eqref{eq:zdeltascaling} for their respective variables. Note that these normalizations are defined through scaling of the data only and therefore are linear transformations, i.e., they do not contain offsets/bias terms. This is done such that we can more easily reconstruct the ROLPVM of the form \eqref{eq:rolpvm}. Let us denote the collections of these normalized data sets as 
\begin{equation}
	\bar{\mc{D}}_\mr{\Delta} = \left\lbrace \left(\bar z(k), \bar z_\Delta, \bar u(k), \bar y(k)\right)\right\rbrace_{k=0}^{N-1}.
\end{equation}

\vspace{1mm}
{\bf Training and reconstruction:}
Denote the (similarly parameterized) learned matrix map $M_\Delta(p)$, scheduling-map $\mu$, and offsets $z_\mr{o}$ and $y_\mr{o}$ on the normalized data matrices by $\bar{M}_\Delta(p)$, $\bar{\mu}$, $\bar z_\mr{o}$, and $\bar y_\mr{o}$, respectively. For training, instead of minimizing \eqref{eq:romreducedobjective}, we minimize the following objective:
\begin{equation}\label{eq:romreduceddeltaobjective}
	\scalemath{0.85}{\min \frac{1}{N}\sum_{k=0}^{N-1}\bigg\lVert\!\begin{bmatrix}
		\bar z_\Delta(k)\\\bar y(k)
	\end{bmatrix} - \left(\!\bar M_\Delta(\bar \mu(\bar z(k),\bar u(k)))\begin{bmatrix}
		\bar z(k)\\ \bar u(k)
	\end{bmatrix}\!+\!\begin{bmatrix}
	\bar z_\mr{o}\\ \bar y_\mr{o}
\end{bmatrix}\right)\!\bigg\rVert^2_{2,\hat\Gamma_2},}
\end{equation}
where $(\bar z(k), \bar z_\Delta(k), \bar u(k), \bar y(k))\in\bar{\mc{D}}_\mr{\Delta}$. Similar to how we learned $\Lambda_\mr{x}$ and $\Lambda_\mr{z}$ for the state-projection in Section \ref{sec:stateproj}, we also use a SGD method to learn $\bar{M}_\Delta(p)$, $\bar{\mu}$, $\bar z_\mr{o}$, and $\bar y_\mr{o}$ to minimize \eqref{eq:romreduceddeltaobjective} to learn the ROLPVM \eqref{eq:rolpvm}. Note that due to the normalization, often there is no need to consider additional weighting in the cost function, i.e., $\hat\Gamma_2=I$ is a reasonable choice.

After learning $\bar{M}_\Delta$, $\bar{\mu}$, $\bar z_\mr{o}$, and $\bar y_\mr{o}$, we can then reconstruct our ROLPVM of the form \eqref{eq:rolpvm} by using  \eqref{eq:mdeltarelation} and the following relations between $\bar{M}_\Delta$ and $M_\Delta$:
\begin{equation}\label{eq:Mdeltanorm}
	M_\Delta(p) = \begin{bmatrix}
		N_\mr{z_\Delta} & 0\\0 & N_\mr{y}
	\end{bmatrix}^{-1} \bar{M}_\Delta(p) \begin{bmatrix}
		N_\mr{z} & 0 \\0 & N_\mr{u}
	\end{bmatrix},
\end{equation}
between $\bar z_\mr{o}$, $\bar y_\mr{o}$ and $z_\mr{o}$, $y_\mr{o}$
\begin{equation}\label{eq:baroffsets}
	\begin{bmatrix}
	z_\mr{o}\\ y_\mr{o}
\end{bmatrix} = \begin{bmatrix}
		N_\mr{z_\Delta} & 0\\0 & N_\mr{y}
	\end{bmatrix}^{-1} \begin{bmatrix}
	\bar z_\mr{o}\\ \bar y_\mr{o}
\end{bmatrix},
\end{equation}
and between $\bar\mu$ and $\mu$:
\begin{equation}\label{eq:schedmapnorm}
	\mu(z,u) = \bar{\mu}(N_\mr{z} z, N_\mr{u} u).
\end{equation}

\section{Examples}\label{sec:examples}
\subsection{3DOF Control Moment Gyroscope}\label{sec:cmg}
In the first example, we apply our approach to learn a ROLPVM of a 3DOF \emph{Control Moment Gyroscope} (CMG). The nonlinear dynamics of a CMG can be derived through Euler-Lagrange equations, see \citep{Bloemers2019} for details, resulting in the differential equation
\begin{equation}
	M(q(t))\ddot{q}(t) + \left(C(q(t),\dot{q}(t)) + F_\mr{v}\right)\dot{q}(t)=K_\mr{m} i(t),
\end{equation}
where $t\in\reals$ is time, $q=\col(q_1,q_2,q_3,q_4)$ are the angles of the four gimbals of the CMG, $i = \col(i_1,i_2,i_3,i_4)$ are the input currents to the motors connected to the gimbals, $M$ is the inertia matrix, $C$ the Coriolis matrix, $F_\mr{v}$ the viscous friction matrix, and $K_\mr{m}$ the motor constant matrix. For the CMG, we consider the case when the gimbal corresponding to $q_3$ is locked and only $i_1$ and $i_2$ are actuated. As outputs of our system, we consider $q_2$, $q_4$, and $\dot{q}_1$. Under these considerations, the dynamics of the CMG can be represented by a continuous-time nonlinear state-space model of the form:
\begin{subequations}\label{eq:nlsysctgyro}
	\begin{align}
		\dot{x}(t) &= f(x(t),u(t));\\
		y(t) &= h(x(t), u(t));
	\end{align}
\end{subequations}
where $x(t) = \col(q_2(t),q_4(t),\dot{q}_1(t),\dot{q}_2(t),\dot{q}_4(t))\in\reals^5$, $u(t)=\col(i_1(t),i_2(t))\in\reals^2$, and $y(t)=\col(q_2(t),q_4(t),\linebreak\dot{q}_1(t))\in\reals^3$. We then discretize \eqref{eq:nlsysctgyro} with sampling time $T_\mr{s}=10^{-3}$ s using a fourth-order Runge-Kutta (RK4) method, where the input is assumed to be constant in the sampling period, giving a model of the form \eqref{eq:nlsys}. 

For the data set $\mc{D}$, we simulate the discretized model for
\begin{subequations}\label{eq:gyroinput}
	\begin{align}
		u(k) = \begin{bmatrix} 0.3+ 0.5 \phi_1(k),\\
		\phi_2(k)\end{bmatrix},
	\end{align}
\end{subequations}
where $\phi_1(k) = \sin(0.1 T_\mr{s} k -0.3 ) + \sin(T_\mr{s} k)+ \sin(10 T_\mr{s} k - 2.1 )$ and $\phi_2(k) = \sin(2 \pi T_\mr{s} k - \pi) + \sin(0.2\pi T_\mr{s}k) + \sin(20\pi T_\mr{s}k-0.1)$, starting from the initial condition $x(0) = \col(0,0,40,0,0)$ for 10001 time instances, i.e., $\sim$10 s. We consider process noise $w$ with a variance $\Sigma_\mr{w} = \diag(0, 0, 9\sn{-4}, 0, 0)$, corresponding to a \emph{Signal-to-Noise-Ratio} (SNR) of 35 dB, and a measurement noise $e$ with a variance $\Sigma_\mr{e} = \diag(8.2\sn{-7}, 2.1\sn{-6}, 1.5)$, also corresponding to an SNR of 35 dB. As our data set $\mc{D}$, we take a random selection of 80\% of this simulation data, i.e., $N = 8000$. An additional data set for the same input trajectory \eqref{eq:gyroinput} is generated \emph{without} any process and measurement noise, which will be used for validation after learning our ROLPVM, see also Fig. \ref{fig:gyro_output}.

For learning of the state-projections $\psi$ and $\psi^\dagger$, we consider a reduced-order state size of $n_\mr{z} = 3$ (while $n_\mr{x}=5$). For $\Lambda_\mr{x}$ and $\Lambda_\mr{z}$, we consider in both cases a single hidden layer with a width of 10 neurons and $\tanh$ activation functions. Furthermore, we consider $\Gamma_1 = \sqrt{N_\mr{x}}$. 
After training the state-projections, we obtain a \emph{Root-Mean-Square Error} (RMSE) error of $6.46\sn{-2}$ between the state and the reconstructed state, i.e., $x-\psi^\dagger(\psi(x))$, on the noiseless simulation data, which was not used for training. Using linear PCA on the same data obtains an RMSE error of $1.55\sn{-1}$. Consequently, our nonlinear state-projection obtains a 58\% improvement compared to linear PCA here. %

For learning of the LPV-NN, we consider a scheduling size of $n_\mr{p}=3$ and 2 hidden layers for the scheduling-map $\mu$, both with a width of 3 $\tanh$ activation functions based neurons. Moreover, based on the nonlinear model, we know that $f(0,0)=0$ and $h(0,0)=0$, therefore, we fix $z_\mr{o}$ and $y_\mr{o}$ to zero and $\hat{\Gamma}_2=I$. %

After training the LPV-NN and constructing the corresponding ROLPVM, we simulate the learned ROLPVM for the input trajectory \eqref{eq:gyroinput} and for the initial condition $z(0) = \psi(x(0)) = \psi(\col(0,0,40,0,0))$ and compare it with the obtained noiseless data set. Moreover, after simulating the reduced-order state trajectory $z$, we reconstruct the full-order state $x$ by $x=\psi^\dagger(z)$. The output trajectory of the ROLPVM of this simulation is depicted in Fig. \ref{fig:gyro_output}. The RMSE and the \emph{Best Fit Rate} (BFR)\footnote{$\text{BFR} = \max\left\lbrace 1 - \sqrt{\tfrac{\sum_{k=0}^{N-1}\norm{v(k)-\hat{v}(k)}_2^2}{\sum_{k=0}^{N-1}\norm{v(k)-v_\mr{mean}}_2^2}},0\right\rbrace\cdot 100\%$, where $v$ is the data sequence, $v_\mr{mean}$ is the sample mean of $v$, and $\hat{v}$ is the predicted response of the model}. of the simulated ROLPVM w.r.t. the validation data are given in Table \ref{tbl:gyro_error}.

From the results, it can be seen that our learned ROLPVM can accurately model the dynamics of the nonlinear CMG on the data set $\mc{D}$. Only output $y_2$ shows slightly higher errors compared to the other outputs. This is due to the state-projection not being able to fully reconstruct the state corresponding to this output for $n_\mr{z}=3$. The full implementation for this and the next example is available at: \url{https://gitlab.com/patrickkoelewijn/sysid-2024-rom-lpv-learning}

\begin{figure}
	\centering
	\includegraphics{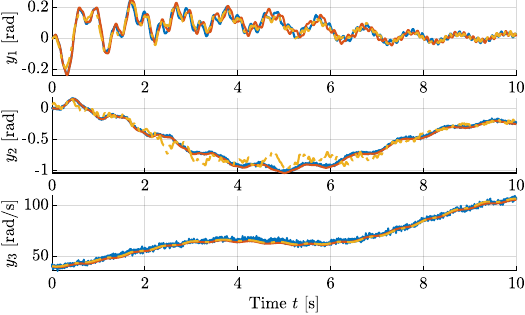}
	\vspace{-2em}
	\caption{Noise infected output of the CMG (\legendline{mblue}), noiseless output (\legendline{morange}), and learned ROLPVM (\legendline{myellow,dashdotted}).}
	\label{fig:gyro_output}
\end{figure}

\begin{table}[t]
\centering
\vspace{-0.1em}
\caption{Error metrics for the CMG ROLPVM.}\label{tbl:gyro_error}
\vspace{-0.3em}
\begin{tabular}{*3l}\toprule
& State ($x$) & Output ($y$) \\ \midrule
RMSE:  & $4.22\sn{-1}$ & $4.70\sn{-1}$\\
BFR: & $94.6\%$ & $95.3\%$\\\bottomrule
\end{tabular}
\end{table}

\subsection{Interconnected Mass-Spring-Dampers System}
In our next example, we consider the nonlinear model of a large interconnection of \emph{Mass-Spring-Damper} (MSD) systems as is considered in \citep{MathWorks2023}. In \citep{MathWorks2023}, a ROLPVM is also constructed using the method of \citePete, to which we will compare our method. %
	The considered system can be represented by a nonlinear state-space model of the form \eqref{eq:nlsysctgyro}, where $x(t)\in\reals^{200}$, corresponding to an interconnection of 100 masses, with an input force $u(t)\in\reals$ only on the first mass, and where the output $y(t)\in\reals$ of the system is the position of the first mass. We refer the reader to \citep{MathWorks2023} for the details.

\begin{figure}
	\centering
	\includegraphics{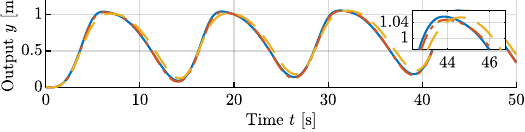}
	\vspace{-2.2em}
	\caption{Output of the MSD (\legendline{mblue}), learned ROLPVM \mbox{(\legendline{morange,dashdotted})}, and result of \citePete (\legendline{myellow,dashed}).}
	\label{fig:msd_output}
\end{figure}

We take the same\footnote{For the implementation of the method of \citePete in \citep{MathWorks2023}, the data is trimmed around equilibrium points, while for our proposed method we take \emph{non-trimmed} responses of the nonlinear system.} data set with $N=2200$ for $\mc{D}$ as the one used in \citep{MathWorks2023}, i.e., a series of chirp input forces with biases. For validation of the ROLPVM, we also use the same data as used in \citep{MathWorks2023}, which is a sinusoidal input force with bias, see also Fig.\ref{fig:msd_output}. For both, no noise is considered, i.e., $w(k)=0$ and $e(k)=0$.

We consider $n_\mr{z} = 5$ (while $n_\mr{x}=200$) for the state-projections $\psi$ and $\psi^\dagger$ and we restrict them to be linear. This means that we consider $\psi(x) = Q^\top x$ and $\psi^\dagger(z) = Q z$ for this example. This is done for the fairness of comparison  against the method in \citePete, which only uses linear state-projection.

We set $n_\mr{p}=3$ for learning of the LPV-NN. Moreover, we consider one hidden layer for the scheduling-map $\psi$ with a width of 10 neurons, we take $\tanh$ activation functions, and take $\hat{\Gamma}_2=I$. Based on knowledge of the nonlinear model of the MSD interconnection, we know that $f(0,0)=0$ and $h(0,0)=0$, therefore, we fix $z_\mr{o}$ and $y_\mr{o}$ to zero. %

After we have trained the LPV-NN and have reconstructed the ROLPVM, we simulate the learned ROLPVM for the input of the validation data set. The simulated output trajectory of our ROLPVM is depicted in Fig. \ref{fig:msd_output}, along with the results of the method of \citePete. The RMSE and the BFR that both methods obtain on the validation data set are given in Table \ref{tbl:msd_error}. Based on these results, it is clear that our method can accurately model the dynamics of the MSD interconnection and our ROLPVM shows a significant improvement in terms of modeling error compared to the method in \citePete. Even though for this example, our proposed method and the method in \citePete both use a linear state-projection, our method can more accurately model the dynamics as it learns a global ROLPVM instead of a local one, which uses interpolation. 

\begin{table}[t]
\vspace{-0.1em}
\centering
\caption{Error metrics of the MSD ROLPVM.}\label{tbl:msd_error}
\begin{tabular}{*5c}\toprule
& \multicolumn{2}{c}{Our method} & \multicolumn{2}{c}{\citePete} \!\!\\ 
 & State ($x$) & Output ($y$) & State ($x$) & Output ($y$) \\\midrule
\!\!RMSE:  & $\bf{1.3\sn{-3}}$ & $\bf{1.3\sn{-3}}$ & ${4.8\sn{-3}}$ & ${5.5\sn{-2}}$\\
\!\!BFR: & $\bf{96\%}$ & $\bf{96\%}$ & $84\%$ & $84\%$\\\bottomrule
\end{tabular}
\end{table}

\section{Conclusions}\label{sec:conclusion}\vspace{-0.3em}
We have presented a novel data-based method for reduced-order LPV modeling of nonlinear systems. In our approach, we first learn a state-projection to a lower dimensional state-space on which we learn an LPV model in the second step using our LPV-NN architecture. Compared to other reduced-order modeling methods, the advantage of our approach is that it learns a structured model in the form of an LPV representation, which allows the application of efficient analysis and controller synthesis methods. Compared to similar LPV based methods, our approach obtains a global reduced order model, instead of a local one, which we have demonstrated in a simulation example to result in an improved representation of the underlying nonlinear dynamics. For future research, we plan to investigate the simultaneous learning of the state reduction and tthe LPV representation.

\nocite{Annoni2017} %
\vspace{-0.2em}

\bibliography{references_short.bib}

\appendix

\end{document}